\magnification=1200
\baselineskip=12pt



\def\bsni{\bigskip\noindent}
\def\msni{\medskip\noindent}
\def\ssni{\smallskip\noindent}
\def\bsbs{\bigskip\bigskip}
\def\ni{\noindent}


\def\a {\alpha} 
\def\b {\beta} 
\def\d {\delta}

\def\l {\lambda}

\def\s {\sigma}

\def\D {\Delta}

\def\O {\Omega}


\def \hlf {{1 \over 2}}
\def \thd {{1 \over 3}} 
\def \qrt {{1 \over 4}}
\def \thqr {{3 \over 4}} 

\def \oln {\overline}
\def \tld {\tilde}

\def \lrar {\longrightarrow}


\centerline{\bf Kinetics of Joint Ordering and Decomposition}

\centerline{\bf in Binary Alloys}

\msni
\centerline{\bf V. I. Gorentsveig,$^1$ P. Fratzl,$^2$ and J. L. Lebowitz$^1$}

\msni 
\centerline{$^1$ {\it Departments of Physics and Mathematics,}}

\centerline{\it Rutgers University, New Brunswick, 08903 New\ Jersey, USA}

\ssni 
\centerline{$^2$ {\it Institut f\"ur Materialphysik,  
Universit\"at Wien, Boltzmanngasse 5, A-1090 Wien, Austria}}

\bsni 
{\bf Abstract}

\msni 
We study phase segregation in a model alloy undergoing both ordering and
decomposition, using computer simulations of Kawasaki exchange dynamics on
a square lattice.  Following a quench into the miscibility gap we observe
an early stage in which ordering develops while the composition remains
almost uniform.  Then decomposition starts with segregation into ordered
and disordered phases.  The two spherically averaged structure functions,
related to decomposition and to ordering, were both observed to obey
scaling rules in the late coarsening stage where the time increase of the
characteristic lengths was consistent with $a(t^{\thd}+b)$.  While $a$ was
similar for ordering and decomposition at low concentration of the minority
component, it showed an increase (decrease) with concentration for ordering
(decomposition).  The domain morphology was found to depend on the
concentration of the minority component, in a way that suggests a wetting
of antiphase boundaries in the ordered domains by the disordered phase.

\vfill \eject 

{\bf 1. Introduction}

\bsni
The kinetics of alloy decomposition is a theoretically challenging and
technologically important problem that has attracted much attention over
many years [1--6].  For alloy systems where the lattice spacings in matrix
and precipitate are the same, and the two-phase structure coarsens with
time to reduce its interfacial energy, both the morphology and the growth
rate of the domains seem to have a universal behavior [7].  In such
systems, the typical domain size grows asymptotically with time $t$ like a
power $t^\alpha$, $\alpha=\thd$ [8, 9], and the essential features of the
morphology depend only on the volume fraction $f$ of precipitates [9, 10].

\ssni
This ``universal'' morphology is known to break down in cases where the
lattice spacings are different in matrix and precipitates. The so-induced
elastic misfit interactions may change the morphology as well as the
kinetics of the decomposition. Such systems have been the subject of
numerous experimental [11--16] and theoretical [17--26] studies in recent
years. It has been found that anisotropic elastic constants lead to the
formation of very anisotropic, mostly plate-like precipitates.Nevertheless,
their typical size still grows in many cases like $t^{\thd}$ [24, 15],
remaining so even under externally applied stresses [25].

\msni
An interesting situation occurs in cases where the precipitates are ordered
intermetallic alloys, such as in Al-Li or in nickel-based superalloys. Here
the decomposition (described by the conserved order parameter $c$, that is
the local concentration of A atoms in the A-B alloy) couples to an ordering
transition (described by a non-conserved order parameter).  In Ni-based
superalloys both the ordering of the precipitates and the elastic misfit
interactions will be present [11--15].  Recent computer work [19, 20, 26,
27] is devoted to the study of models that include both effects.  While
some features of the morphology, like the cuboidal shape of the
precipitates and their predominant alignment along certain crystallographic
directions, are clearly due to the elastic misfit interactions, some other
features are probably due to the ordering of the precipitates. Indeed, it
is usually found in these alloys that close neighboring precipitates do not
merge even though this would reduce the total surface energy [11--15].  A
possible explanation, proposed in [20], is that these precipitates belong
to different variants of the same ordered structure so that their merging
would lead initially to an anti-phase boundary (APB), whose energy is
higher than that of the interfaces between the ordered precipitates and the
disordered matrix.

\msni
In order to address this issue and to gain a better understanding of the
influence of atomic ordering on the coarsening kinetics and morphology, we
have carried out computer simulations of an Ising model on a square lattice
with nearest and next-nearest neighbor interactions, which does not include
elastic misfit interactions.  There will now be two variants of the ordered
phase depending on which sublattice is occupied by which type of atoms and
if a wetting of APB's by the disordered phase reduces the total energy,
then the domain morphology will be very different from the case where none
of the phases (minority or majority) is ordered.  In particular at larger
volume fractions $f$ of the ordered phase, e.g.\ when $f>0.5$, APB's are to
be expected between the differently ordered domains.  It is unclear whether
the length scale of precipitates in such a system will still grow like
$t^{\thd}$ or will behave like $t^{1/2}$, the typical growth law for
nonconserved quantities.

\msni 
The model we investigate was studied previously by Sahni et al. [28].
Their conclusions were mainly qualitative. In the present work we extend
their simulations and obtain more quantitative information about the joint
kinetics of ordering and segregation. We concentrate on comparing the
growth kinetics and morphology at $f<0.5$, $f=0.5$ and $f>0.5$ with those
obtained for an alloy where none of the phases has an ordered intermetallic
structure.  While the $t^{\thd}$ law appears to hold in all cases, the
coefficient $a$ has very different dependence on concentration for the
ordering and segregation, the former growing much faster for higher
concentrations.  We also find, in the case of high volume fraction of
ordered precipitates, clear evidence of the wetting of APB's.  This may
explain some of the results in Al-Li systems [29, 30] where ordered domains
with little lattice mismatch are formed and supports the interpretation of
non-merging precipitates given in [29, 30].  We plan to extend our work to
include elastic interactions, present in Ni-based alloys [11--15],
mentioned earlier.

\bsni\bsni 
{\bf 2. The Model}

\bsni 
We consider an Ising model on a 2D square lattice with nearest neighbor
interactions $J$ of antiferromagnetic type and next nearest neighbor
interactions of ferromagnetic type, which in our simulations we took to be
$-J/2$.  The total energy $U$ of this system, wrapped on a torus containing
$N = L^2$ sites, is
$$U = 
J\sum_{\langle {\bf l} , {{\bf l}'} \rangle_+}
\s_{\bf l}\s_{{\bf l}'} - 
\hlf J\sum_{\langle {\bf l} , {{\bf l}'} \rangle_\times}
\s_{\bf l}\s_{{\bf l}'}, \quad \quad J > 0,  \eqno (1) $$ 

\ni 
where $\s_{\bf l}=1$ if there is an A-atom and $\s_{\bf l}=-1$ if there is
a B-atom, on the lattice site ${\bf l = (l_1,l_2)}$ and the first and the
second sums are over all nearest and next nearest neighbor pairs,
respectively. The average concentration $c$ of minority B-atoms is
$$c = (1-\overline \s)/2 $$ 

\ni 
where 
$$\oln \s = {1 \over N}\sum_{\bf l} \s_{\bf l}.$$  

\ni

\ssni
The phase diagram of this system obtained from Monte-Carlo simulation [31,
32] is shown schematically in Fig. 1.  The critical point for the ordering
transition is at $c_c=0.5$ and $T_c \simeq 3.8\,J/k$ and there is a
tricritical point at $c_t \simeq 0.27$ and $T_t \simeq 1.3\,J/k$.  Above
$T_t$ the transition between the disordered A-rich phase and the ordered
phase is of 2nd order, while below $T_t$ the transition is of the 1st
order. Fig.1 shows the miscibility gap between the ordered phase (with
practically stoichiometric A-B composition) and the disordered phase.

\ssni
Our simulations were performed on a square lattice, $128 \times 128$, with
periodic boundary conditions, at $T=0.9\,J/k$ at three values for the
composition, $c=0.15$, $0.25$ and $0.35$.  These are all well within the
two-phase region of Fig.\ 1 between the ordered intermetallic alloy with
stoichiometric composition AB and the disordered A-rich phase;
corresponding approximately to volume fractions the respective volume
fractions of ordered phase will therefore be about $f=0.3$, 0.5 and 0.7,
respectively.  The dynamics of the ordering/decomposition process, which
are our main concern, were followed using Kawasaki dynamics with the
Metropolis rule, that is by choosing a nearest neighbor pair at random and
then exchanging the atoms with probability
$$w=\min[1,\exp(-\Delta U/kT)]$$

\ni
where $\Delta U$ is the change in the total energy $U$ given by
eq.(1). 

In our simulations we start the realizations (computer runs) from randomly
generated ``infinite'' temperature configurations.  Time is then measured
in Monte-Carlo Steps (MCS), that is the number of attempted nearest
neighbor exchanges per lattice site.

\bsni \bsni
{\bf 3.  Characterization of Configurations}

Since there are two sublattices within the ordered "antiferromagnetic"
structure on the square lattice, there will also be two possible variants
for the ordered phase, one where the A-atoms are on the even sublattice
(which we call for short the even variant) and one where the A-atoms are on
the odd sublattice (odd variant). To represent "snap-shot" pictures of the
configurations in a way that clearly differentiates between the two variant
ordered phases as well as the disordered phase, we use the following
variable,
$$\eta_{\bf l} = \hlf (\s^s_{\bf l} +  
		\qrt \sum_{\langle {\bf l}' \rangle^+_{\bf l}} 
			 \s^s_{{\bf l}'}), \eqno (2)$$

\ni 
where the sum is over nearest neighboring sites, and
$$\s^s_{\bf l} = (-1)^{l_1+l_2} \s_{\bf l} $$

\ni 
is the staggered spin.  The variable $\eta_{\bf l}$ takes the value $-1$ at
sites within ordered domains of the odd variant and +1 on sites within
ordered domains of the even variant. In the disordered phase, where almost
all atoms are of the same kind, $\eta_{\bf l}=0$.  Intermediate values
$\eta_{\bf l}=\pm \qrt,\pm \hlf, \pm \thqr$ appear on the interfaces
between ordered and disordered domains and at isolated B-atoms dissolved in
the A-rich (disordered) phase. Translating this into a grey-scale, the two
variants of the ordered phases appear black and white, respectively, while
the disordered phase appears in grey, see Fig. 2.

The staggered spin local average, $\eta_{\bf l}$, can be interpreted as
local order parameter for ordering and its absolute value averaged over the
lattice,
$$\eta_{sr}\equiv \oln{|\eta|} 
	= {1 \over N}\sum_{\bf l}|\eta_{\bf l}| \eqno (3)$$      

\ni 
can be considered as a short range order parameter (for the atomic
ordering, as opposed to decomposition). It represents the relative amount
of ordered phase, irrespective of whether the order corresponds to the odd
or the even variant.

\ssni
The average over the entire lattice of the staggered spin local
average, 
$$\eta_{lr} \equiv \oln\eta =  {1 \over N}\sum_{\bf l}\eta_{\bf l}  
	= {1 \over N}\sum_{{\bf l}} \s^s_{{\bf l}}, \eqno (4)$$  

\ni 
is equal to the global average of the staggered spin variable and can be
considered as a long range order parameter. It corresponds to twice the
difference in the concentration of A-atoms between the odd and the even
sublattice. Similar order parameters may also be defined for decomposition,
although their interpretation is more ambiguous now,
$$\rho_{sr}\equiv {1 \over N}\sum_{\bf l}|\rho_{\bf l}|, 
\qquad \rho_{lr} \equiv {1 \over N}\sum_{\bf l}\rho_{\bf l} = 1 - 2 c 
\eqno (5)$$      

\ni
where
$$\rho_{\bf l} = \hlf (\s_{\bf l} +  
		\qrt \sum_{\langle {\bf l}' \rangle^+_{\bf l}}  
			 \s_{{\bf l}'}).$$

\ni
{}From these definitions it is apparent that the long-range order parameter
$\rho_{lr}$ for decomposition is conserved (since $c$ is fixed) while
$\eta_{lr}$ is non-conserved. One may further notice that an alternative
way to write $\eta_{sr}$ is
$$ \eta_{sr}  = 1-\rho_{sr} =
\hlf-{1 \over {4N}}  
\sum_{\langle {\bf l} , {{\bf l}'} \rangle_+}  \s_{\bf l} \s_{{\bf l}'} 
= 2\,c\,(1-c)\,(1-\alpha_1). \eqno (6)$$

\ni
This expression shows that $\eta_{sr}$ is related linearly to the first
Warren-Cowley short-range order parameter $\a_1$ generally used in
crystallography [33],
$$ \alpha_1=1-p^{AB}\left/ c \right. ,$$

\ni
where $p^{AB}$ is the density of B-atoms at nearest neighbor sites of an
A-atom.
 
\msni
To characterize the decomposition kinetics, we use the structure function
$$S_{{\bf k}} = {1 \over N}|\sum_{\bf l}
		{\rm exp}(i{\bf k} \cdot {\bf l}) \hskip 1pt \s_{\bf l}|^2,
		\eqno (7)$$ 

\ni 
where ${\bf k}=(k_1,k_2)$ is a reciprocal lattice vector, with $k_j=2\pi
K_j/L$ and $K_j=1,2,...,L$ for $j=1,2$.  For the system undergoing both
ordering and decomposition, the structure function (shown in Fig. 3 by
means of brightness) is localizing in two areas, around the point
$(\pi,\pi)$ (related to ordering) and around the point $(0,0)$ (related to
decomposition) of reciprocal space. Note, that $S_{(0,0)}=N\rho_{lr}^2$ and
that $S_{(\pi,\pi)}=N\eta_{lr}^2$.

\ssni 
To study kinetics and scaling properties of the two processes we divide the
(square shaped) Brillouin zone in reciprocal space in two equal areas (bold
lines in Fig. 3) : the square for ordering,
$$\O_\pi = \{ {\bf k}: |k_1-\pi| + |k_2-\pi| \leq \pi \}
\eqno (8)$$ 

\ni 
(that is inside that boundary in Fig. 3) and the remaining four quarters of
the square for decomposition. Using the periodicity of the reciprocal
lattice, the latter can be put together as a square,
$$\O_0 = \{ {\bf k}: |k_1| + |k_2| \leq \pi \} \eqno (9)$$ 

\ni 
shown separately in Fig. 3 with higher resolution for values of the
structure function.  To test scaling behavior of the structure function, we
sphericalize it within the two squares $\O_\nu$ related to ordering
($\nu=\pi$) and to decomposition ($\nu=0$), separately. We define:
$$S_\nu (k) = \sum_{\D_\nu(k)} S_{{\bf k}'} / 
	     \sum_{\D_\nu(k)} 1, \eqno (10)$$

\ni 
where
$$\D_\nu(k)=\{ {\bf k}': k - \d k < | {\bf k}' - (\nu,\nu)| 
	    \leq k + \d k\}$$  

\ni 
is a ring of width $2\d k$ and median radius $k=|{\bf k}| =(k^2_1 +
k^2_2)^\hlf $, centered at $(\nu,\nu)$.  The sphericalized structure
function was actually computed for the values $k = 2\pi K/L$, $K$ being an
integer: $K=1,2,...,[L/2\surd \oln 2]$, and choosing $\d k=\pi/L$. Since
the configurations on the lattice are evolving with time, we are also using
the notation $S_\nu(k,t)$ for the sphericalized structure function computed
for the configuration at time $t$.

The scaling hypothesis for the late stages of coarsening, in systems
undergoing either phase segregation or ordering, is that the system is
characterized by a single scale $\lambda(t)$.  This means in particular
that for values of $k^{-1}$ which are large compared to the lattice spacing
and small compared to the size of the system $S(k,t)$ should (up to a time
dependent factor) be a function of only one variable, $\lambda(t) k$.  In
our case we need a priory two characteristic lengths, $\lambda_\nu (t)$,
$\nu = 0, \pi$.  We therefore hypothesize that for the system under
consideration
$$\tld S_\nu (k,t) \approx  A_\nu (t) \;  F_\nu (k\lambda_\nu (t)) 
\hskip 12pt {\rm as} \hskip 12pt t \lrar \infty, \eqno (11)$$ 

\ni
where $\tld S_\nu (k,t)$ is the macroscopic, formally the infinite volume
limit, for fixed $k \ne 0$, of the corresponding sphericalized structure
function at time $t$ [9]:
$$S_\nu (k,t) \lrar \tld S_\nu (k,t) \hskip 12pt 
{\rm as} \hskip 12pt L \lrar \infty$$ 
We shall show, in section 4, that this generalized scaling hypothesis is
indeed consistent with our simulation data.  

One of the common ways of defining a characteristic length is to use
inverse moments of the structure function.  Since we are restricted in
simulations to systems which are very small on the macroscopic scale, great
care has to be taken in computing from the available data quantities
relevant for macroscopic systems.  This is doubly so for our system where
we have both ordering and segregation.  We therefore had to find a way
which minimizes distortions of the time dependence of the $\lambda_\nu$ due
to the finite size of the system.  To achieve this we defined the moments

$$k_\nu(t) = \sum_{k_a}^{k_b} k^{1+r_\nu} S_\nu (k,t)  /
\sum_{k_a}^{k_b} k^{r_\nu} S_\nu (k,t) \eqno (12)$$

\ni
with $r_0=1$ and $r_\pi=2$, where the limits of summation, $k_a$ and $k_b$,
were themselves scaled in accordance with $k_\nu(t)$, by putting
$k_{a,b}(t) = (1 \pm h) k_\nu(t);$ so at different times the moments are
calculated from similar intervals.  Taking $h = .7$, the interval around
$k_\nu$ was large enough to give sufficient statistics for the summation,
but still small enough to avoid any contribution from the structure
function at very small or very large $k$, which might be affected by finite
size effects.  The $k_\nu(t)$ so obtained appeared to follow the position
of the maximum of $k^{r_\nu} S_\nu(k,t)$ in eq. (12).

With the procedure described above, finite size effects on $k_\nu(t)$ could
be minimized. Indeed, additional runs performed on a smaller lattice
(64x64) gave consistent results for the time evolution of $k_\nu(t)$.  Of
course this still leaves considerable uncertainties in determining $k_\nu
(t)$.  For the size of system we are dealing with, the gaps in the $k$'s,
which are equal to $2\pi/L$, limit any finer resolution independent of the
number or duration of the runs.

\bsni\bsni 
{\bf 4. Results} 
 
\bsni
Snapshot pictures of typical configurations obtained at $T=0.9J/k$ for
alloy compositions $c=0.15$, 0.25 and 0.35 with lattice size $L=128$ are
shown in Fig. 2. The value for the locally averaged spin $\eta_{\bf l}$,
defined in eq.(2) is shown at 125, 8000 and 64000 MCS. Three different
types of domains appear in these pictures in white, black and grey
corresponding to ordered phases on the even and the odd sublattice and the
disordered A-rich phase, respectively. In the case $c=0.25$, the grey phase
covers about half of the specimen volume, while at $c=0.35$ it covers less
and at $c=0.15$ more. This is, as already mentioned, in qualitative
agreement with the equilibrium phase diagram [28, 31, 32] where the
predicted volume fractions of black or white (that is ordered) phase are
$f=0.3$, 0.5 and 0.7 for $c=0.15$, 0.25 and 0.35, respectively, see Fig. 1.
In the case of decomposition without ordering, when the miscibility gap is
symmetric around the critical composition, the behavior of an alloy with
$f=0.3$ would be identical to the one at $f=0.7$, with the two phases
exchanged in their respective role. In the present case, however, it
becomes clear from looking at Fig. 2 that there is a completely different
behavior on different sides of the tricritical composition.  At $c = .15$
the pictures show isolated ordered droplets within a disordered matrix,
while for $c = .35$ they show the disordered phase surrounding the ordered
domains, i.e. we see an ordered alloy with the antiphase boundaries between
the two ordered variants wetted by the disordered phase.

\bsni 
{\bf 4.1. Early Stage Behavior}

\bsni 

We observe that the short range order parameter, $\eta_{sr}$ comes rapidly
to the steady value $2c$, corresponding to complete ordering. We find that
the relative deviation becomes small, $|\eta_{sr}/2c - 1|
\leq 10^{-2}$, by the time $t \approx 10^2$,  indicating that local
ordering is well developed by that time.  At the same time a domain
structure starts to be visible (see Fig. 2).  Ordered domains of both kinds
(even and odd variant) occur in about equal fractions. Domains of pure
minority component are practically not observed (in the original spin
configuration) after time $t \sim 10^2$. Then segregation of ordered and
majority phases develops.

\bsni 

We further observe that the sphericalized structure function for ordering,
$S_\pi (k)$, develops a maximum localized at $k \sim 0$ by the time $t \sim
10$ (starting from times $t \sim 1$) which grows sharper as $t$ increases
(Fig. 4.1), while the sphericalized structure function for decomposition,
$S_0 (k)$, develops a maximum at $k \sim k_0$ later, starting at $t \sim
10^2$ (Fig. 4.2). This shows, in agreement with what was found in [29, 30,
34], that, initially, ordering is much faster than decomposition.

\bsni 
Moreover, the sum of the structure function components related to ordering
$$ N\,s_\pi = \sum_{\O_\pi} S_{\bf k} \eqno (13) $$

\ni
approaches the constant value $2Nc$, which corresponds to the area of
ordered phase at complete phase separation. Similarly, the sum of the
structure function components related to decomposition (that is, within the
square $\O_0$ with exclusion of the origin),
$$N \, s_0 = \sum_{\O'_0} S_{{\bf k}} \eqno (14) $$ 

\ni
also reaches a steady value $N s_{0}^c$, with $s_{0}^c \simeq (1-2c)2c$,
which was to be expected, since there is a general relation between $s_0$
and $s_{\pi}$:
$$s_\pi + s_0 =1- S_{(0,0)}/N=4\,c\,(1-c). \eqno (15) $$      

\ni
The relative deviation of $s_0$ from its equilibrium value, $|s_{0}/s_{0}^c
- 1|$, becomes smaller than $10^{-2}$, for times $t > 10^2$.

\bsni 
{\bf 4.2. Scaling Analysis} 

\bsni

As already mentioned, both spherical averages of the structure functions,
$S_\pi (k)$, and $S_0 (k)$ defined in (10), have a well defined single
maximum by the time $t \sim 10^3 \ or \ 10^4$ (see Figs. 4.a and 4.b).
Using these functions to compute the moments defined in (12) we find that
the two characteristic dimensions related to ordering and decomposition,
$\l_\nu(t)$, satisfy [35, 9] 
$$\l_\nu(t) = 2\pi/k_\nu(t) \approx a_\nu (t^{\thd} + b_\nu) 
\eqno (16) $$

\ni 
for times greater than 10$^3$. This can be seen in Figs. 5.a and 5.b, for
ordering and decomposition, respectively. The constants $a_\nu$ and $b_\nu$
are given in Table 1.  Equation (16) shows that the growth of all
characteristic dimensions is consistent with the fimiliar $t^{1/3}$ growth
law [8, 35] although an unambiguous determination of the growth law
exponent would only be possible with much longer runs on a considerably
larger lattice.

We now investigate the scaling hypothesis for both functions which requires
that $S_\nu(k,t)$ can be written in the form

$$S_\nu(k,t) \approx B_\nu \, k'_\nu(t)^{-2} \,F_\nu(k/k'_\nu(t)),
\eqno (17)$$ 

\ni
where we have used the constancy of the sums in (13) and (14); $k'_\nu(t)$
may differ from the previously defined $k_\nu(t)$ by a time-independent
(but composition-dependent) factor, i.e.

$$ 2\pi/k'_\nu(t) = a'_\nu \,(t^\thd+b_\nu). \eqno (18)$$

\msni
A standard way to choose the constants $a'_0$ and $B_0$ is to require that
the scaling function $F_0$ satisfy the normalization condition
$${\rm max}[  F_0(x)] =  F_0(1) = 1. \eqno (19) $$        
This permits comparison of characteristic parameters at different
concentrations.  

To better locate the maximum of the function $F_0(x)$, we tried
various empirical fits starting with a one parameter fitting
expression proposed for decomposition in two dimensions, Eq. (14) in
[9].  While this works well for pure segregation and gave a reasonable
fit to our data, it appeared to miss the stronger asymmetry, about the
maximum, present in our system.  This was particularly apparent at $c
= .35$ when the morphology is indeed very different from the pure
segregation case.  We found by trial and error that the following
formula
$$
F_0(x) = {\alpha x^4/(\gamma + x^4)}[\beta + (x^2 - 1 + \delta)^2]^r,
\quad \quad \alpha, \beta, \gamma > 0, \eqno(21)
$$
$r = 3/4$, gave a good fit.  The form (21) with $3/4$ replaced by $(d+1)/4$
is a generalization, satisfying Porod's law $(F_0(x) \sim x^{-(d+1)} {\rm
at} \quad x \to \infty)$ [36], of a heuristic formula suggested in [10] for
3D alloys (where $d = 3$).  Definition of the standard form function
$F_0(x)$ yields two relations between parameters
$$
\alpha = (1 + \gamma)(\beta + \delta^2)^r, \quad \quad \gamma =
r\delta/(\beta - r\delta + \delta^2)
$$
so, two of the four parameters, preferably $\beta$ and $\delta$, are
independent.  These were fitted from the data.

For $F_\pi$ the situation was more difficult, because there was no scaling
close to $(\pi,\pi)$.  The reason for this is that $S(\pi,\pi)$ is related
to the (non-conserved) long-range order parameter
($S(\pi,\pi)=N\,\eta_{lr}^2$), which depends strongly on how the ordered
phase is distributed over the two sublattices.  Further away from the
$(\pi,\pi)$-point, the structure function describes the short-range order,
which is not sensitive to the difference between the two
sublattices. Because of that, $a'_\pi$ and $B_\pi$ were determined by
fitting of $F_\pi(x)$ in the region where scaling holds, to the square of
the Lorentzian function,
$$F_\pi(x) = 1/(1 + x^2)^2,  \eqno (22)$$

As shown in Fig.6, we observe scaling of both structure functions (for
ordering and for decomposition) for times $t \sim 10^4$ through $t \sim
10^5$.  The obtained values for the constants $\beta$, $\delta$, $a'_\nu$
and $B_\nu$ are given in Table 2.  We find that $a'_\pi$, giving the rate
of ordering kinetics, increases while the rate $a'_0$ of decomposition
(coarsening) decreases with the amount of ordered phase at equilibrium.
This is presumably due to the very different morphologies at high and low
concentrations discussed earlier.

Finally, the dependence of full-width at half maximum (FWHM) of the scaling
function on the volume fraction of B-rich phase is given in Fig. 7 where it
is compared with the $\rm FWHM_{\rm ferro}$ obtained from simulations on
the two-dimensional Ising model with attractive nearest neighbor
interaction between like atoms [9].  The $\rm FWHM_{\rm ferro}$ is
practically temperature independent (for $0.34 \leq T/T_c \leq 0.8$, $T_c$
being the critical temperature) and symmetric around the volume fraction
$f=0.5$. In the present model, however, the scaling function is wider in
all cases, indicating a smaller amount of regularity in the positions of
the precipitates. Moreover, there is a clear asymmetry between $f\approx
0.3$ and $f \approx 0.7$, which is not surprising since the morphologies of
the two-phase mixtures are also quite different at these compositions
(cf. Fig. 2).

\bsbs
\bsni
{\bf Discussion} 

\bsni 
The general picture emerging for the ordering/decomposition process in the
present model is that the phase transformation starts with the development
of (short-range) order followed by the appearance of well-ordered domains
inside a disordered matrix.  At this stage, the ordering (as characterized
by the short-range order parameter $\eta_{sr}$, eq.3) has reached its
maximum value and the further development corresponds to the coarsening of
domains by long-range diffusion, much like in a situation where
decomposition into two pure (or disordered) phases occurs. The fact that
ordering appears prior to decomposition has been observed experimentally
[29, 30, 11--15] and can be explained by the fact that the diffusion of
B-atoms is required only over short distances for the development of atomic
order but needs to cover long distances for the decomposition into phases
with different amounts of B-atoms.

\msni
The coarsening kinetics of the domains, measured by the correlation lengths
for decomposition, $k_0^{-1}$, and for ordering, $k_\pi^{-1}$, is
consistent with a growth of domain size proportional to $t^{1/3}$, in
agreement with experiments on ordered precipitates [29, 39, 11--15]. This
growth law, as well as the observed time-scaling of the structure function,
are a signature of the coarsening of two-phase structures under the
influence of interfacial energy [9, 35, 37]. The most striking difference
between the present case and a conventional coarsening process (with two
disordered or pure phases) is the composition dependence of the morphology.

In particular, the lack in symmetry between the case $c = 0.15$ (with
volume fraction $f \approx 0.3$) and $c = 0.35$ (with volume fraction $f
\approx 0.7$) may be related to the presence of ordered domains of two
kinds (on the two sublattices) with some additional positive energy
appearing (as pointed out, e.g., in [29, 30, 34]) on their interface when
they are coming in contact due to their growth (anti-phase boundary).
These anti-phase boundaries (APB's) appear to be wetted by the disordered
phase which reduces the overall energy of the system. Hence, even though
the ordered phase is the majority at $c=0.35$, individual ordered droplets
do not join up but stay separated by narrow channels of disordered phase
(see Fig.2c).  The composition-dependent morphology may also be at the
origin of the dramatically different behavior of the typical length scales
for ordering $\lambda_\pi$ and for decomposition $\lambda_0$, when the
alloy composition is changed.
\bigskip
\noindent
{\bf Acknowledgments}

We thank Sorin Bastea, Claude Laberge, Armen Khachaturyan and Yunzhi Wang
for their help.  Work supported in part by NSF Grant NSF--DMR 92--13424,
and FWF Grant S5601.
\vfill \eject

{\bf References} 

\ssni
[1] J.W. Cahn, {\it Acta Metall.} {\bf 9}, 795 (1961); {\bf 10}, 179 (1962).

\ssni
[2] A.B. Bortz, M.H. Kalos, J.L. Lebowitz and M.H. Zendejas, {\it Phys. Rev. B}
{\bf 10}, 535 (1974); J. Marro, A.B. Bortz, M.H.Kalos and J.L. Lebowitz,
{\it Phys. Rev. B} {\bf 12}, 2000 (1975).

\ssni
[3] J.D. Gunton, M. San Miguel and P.S. Sahni, in  {\it Phase Transitions 
and Critical Phenomena}, edited by C. Domb and J.L. Lebowitz (Academic Press, 
New York, 1983), Vol.~8.
 
\ssni
[4] K. Binder, in  {\it Materials Science 
and Technology}, edited by P. Haasen (VCH, Weinheim, 1991), Vol.~5, Chap.~7.

[5] R. Wagner, R. Kampmann, in {\it Phase Transformations in 
Materials}, edited by P. Haasen (VCH, Weinheim, 1991), Vol. 5, Chap. 4.

\ssni 
[6] A.G. Khachaturyan, 
{\it Theory of Structural Transformations
in Solids} (Wiley, New York, 1983). 

\ssni
[7] J. Marro, J.L. Lebowitz and M.H. Kalos, 
{\it Phys. Rev. Lett.} {\bf 43}, 282 (1979); J.L. Lebowitz, J. Marro
and M.H. Kalos, {\it Acta Metall.} {\bf 30}, 297 (1982); P. Fratzl,
J.L. Lebowitz, J. Marro and M.H. Kalos, {\it Acta Metall.} {\bf 31},
1849 (1983).

\ssni
[8] I.M. Lifshitz and V.V. Slyozov, {\it J. Phys. Chem. Solids}
{\bf 19}, 35 (1961); C. Wagner, {\it Z. Elektrochem.} {\bf 65}, 568 (1961).

\ssni
[9] P. Fratzl, J.L. Lebowitz, 
O. Penrose, J. Amar, {\it Phys. Rev. B} {\bf 44}, 4794 (1991).

\ssni
[10] P. Fratzl and J.L. Lebowitz, {\it Acta Metall.} {\bf 37}, 3245 (1989).

\ssni
[11] T. Miyazaki, M. Doi and T. Kozaki, {\it Solid State Phen.}
{\bf 3} {158} (1988); T. Miyazaki and M. Doi, {\it Mater. Sci. Eng.} 
{\bf A110}, 175 (1989).

\ssni
[12] A. Maheshwari and A.J. Ardell, {\it Phys. Rev. Lett.}
{\bf 70}, 2305 (1993).

\ssni
[13] O. Paris, M. F\"ahrmann and P. Fratzl, {\it Phys. Rev. Lett.}
{\bf 75}, 3458 (1995).

\ssni
[14] A.D. Sequeira, H.A. Calderon, G. Kostorz and J.A. Pedersen,
{\it Acta Metall. Mater.} {\bf 43}, 3427 (1995); {\bf 43}, 3441 (1995).

\ssni
[15] M. F\"ahrmann, P. Fratzl, O. Paris, E. F\"ahrmann and
W.C. Johnson, {\it Acta Metall. Mater.}, 1007 (1995).

\ssni
[16] P. Fratzl, F. Langmayr, G. Vogl and W. Miekeley, {\it Acta 
Metall. Mater.} {\bf 39}, 753 (1991);
F. Langmayr, P.Fratzl, G.Vogl and W. Miekeley, 
{\it Phys. Rev. B}, {\bf 49}, 11759 (1994).

\ssni
[17] J.D. Eshelby,  {\it Prog. Solid Mech.} {\bf 2}, 89 (1961);
Ardell, R.B. Nicholson and J.D. Eshelby, {\it Acta metall.} 
{\bf 14} {1295} (1966).

\ssni
[18] A. Onuki and H. Nishimori, {\it Phys.Rev.B} {\bf 43}, 
13649 (1991); H. Nishimori and A. Onuki, {\it Phys. Rev. B}
{\bf 42}, 980 (1990).

\ssni
[19] Y. Wang, L.Q. Chen and A.G. Khachaturyan, 
{\it Acta Metall. Mater.} {\bf 41}, 279 (1993);
Y. Wang and A.G. Khachaturyan, {\it Acta Metall. Mater.} {\bf 43},
1837 (1995);

\ssni
[20] Y. Wang and A.G. Khachaturyan, {\it Scripta Metall. Mater.}
{\bf 31}, 1425 (1994).

\ssni 
[21] W.C. Johnson, {\it Acta metall.} {\bf 32}, 465 (1984);
T.A. Abinandanan and W.C. Johnson, {\it Acta Metall.
Mater.} {\bf 41}, 17 (1993).

\ssni 
[22] P.W. Voorhees, G.B. McFadden and W.C. Johnson,
{\it Acta Metall. Mater.} {\bf 40}, 2979 (1992).

\ssni
[23] K. Kawasaki and Y. Enomoto,  {\it Physica} {\bf A150}, 463 
(1988); Y. Enomoto and K. Kawasaki,  {\it Acta Metall.} {\bf 37}, 1399 (1989).

\ssni
[24] P. Fratzl and O. Penrose, {\it Acta Metall. Mater.} {\bf 43}, 2921 
(1995).

\ssni
[25] C.L. Laberge, P. Fratzl and J.L. Lebowitz, {\it Phys. Rev. Lett.}
{\bf 75}, 4448 (1995).

\ssni
[26] W.A. Soffa and D.E.Laughlin, {\it Acta Metall.} {\bf 37}, 3019 (1989).

\ssni
[27] C. Sagui, A.M. Somoza and R.C. Desai, {\it Phys. Rev. E} {\bf 50},
4865 (1994).

\ssni
[28]  P.S.Sahni, J.D.Gunton, S.L.Katz, and R.H.Timpe, 
  {\it Phys. Rev. B} {\bf 25}, 389 (1982).

\ssni
[29] O. Blaschko, R. Glas, and P. Weinzierl, {\it Acta Metall. Mater.} 
{\bf 38}, 1053 (1990).

\ssni
[30] H. Okuda, V. Vezin, K. Osamura, and Y. Amemyia,
{\it Proc. Solid $\rightarrow$ Solid Phase Transformations}, p. 371
ed. W.C. Johnson et al., The Minerals, Metals \& Materials Society, 1994.

\ssni
[31] D.P.Landau, {\it J.Appl.Phys.} {\bf 42}, 1284 (1971). 

\ssni
[32] D.P.Landau, {\it Phys. Rev. Lett.} {\bf 28}, 449 (1972).

\ssni
[33] D. de Fontaine, {\it Solid State Physics} {\bf 34}, 74 (1979).

\ssni
[34] L.-Q. Chen and A.G. Khachaturyan, {\it Acta Metall. Mater.} 
{\bf 39}, 2533 (1991). 

\ssni
[35] D.A. Huse, {\it Phys. Rev. B} {\bf 34}, 7845 (1986). 

\ssni
[36] G. Porod, {\it Kolloid-Zeitschrift} {\bf 124}, 83 (1951); 
{\bf 125}, 51 (1952).

\ssni
[37] N. Akaiwa and P.W. Voorhees, {\it Phys. Rev. E} {\bf 49}, 3860 (1994).

\ssni
[38] P. Fratzl and O.Penrose, {\it Acta Mater.} {\bf 44}, 3227 (1996).
\vfill \eject

\ni
Table 1. Parameters $a_\nu$ and $b_\nu$ characterizing the growth (16) of
of the characteristic lengths defined in (12), at different concentrations
$c$ of minority component.

\baselineskip=12pt
\bigskip\bigskip\ni 
$ \hskip 6pt  
   c   \hskip 30pt  a_\pi  \hskip 24pt  b_\pi  \hskip 18pt
       \hskip 24pt  a_0    \hskip 18pt  b_0        $

\ni
---------------------------------------------------------------------------

\ni
$  0.15 \hskip 18pt 0.78   \hskip 20pt  9.0  \hskip 20pt 
       \hskip 15pt 0.77   \hskip 18pt  7.0     $

\ni
--------------------------------------------------------------------------- 

\ni
$  0.25 \hskip 18pt 0.94    \hskip 15pt  11.0    \hskip 20pt
       \hskip 15pt 0.67    \hskip 18pt   5.5       $

\ni
---------------------------------------------------------------------------

\ni
$  0.35 \hskip 18pt  1.7    \hskip 21pt    10.0   \hskip 20pt 
       \hskip 15pt   0.475  \hskip 13pt     9.0      $

\ni
--------------------------------------------------------------------------- 

\baselineskip=24pt
\vskip 1in
\ni 
Table 2. Parameters $a'_\nu$ and $B_\nu$ used for the scaling of the
structure functions, eqs.(17) and (18), and parameters $\b$ and $\delta$
used for the fitting of $F_0(x)$ with eq.(21), as a function of the
concentration of minority atoms, $c$.

\baselineskip=12pt
\bigskip\bigskip\ni 
$ \hskip 6pt 
 c   \hskip 30pt a'_\pi  \hskip 18pt B_\pi  \hskip 18pt
     \hskip 20pt a'_0    \hskip 18pt B_0   
     \hskip 30pt \b \hskip 30pt \delta  $

\ni
\vskip.1truein
\hrule  
\vskip.1truein
\ni
$  .15 \hskip 23pt  0.82   \hskip 15pt  3.1   \hskip 20pt     
       \hskip 15pt  0.97   \hskip 15pt   0.53   
       \hskip 18pt    1.47   \hskip 20pt 0.13 $ 

\ni
\vskip.1truein
\hrule  
\vskip.1truein
\ni
$  .25 \hskip 23pt 1.01   \hskip 15pt  5.1    \hskip 20pt  
       \hskip 15pt 0.79   \hskip 15pt  0.75    
       \hskip 18pt 0.98   \hskip 20pt  0.065    $

\ni
\vskip.1truein
\hrule  
\vskip.1truein
\ni
$  .35 \hskip 23pt 1.75   \hskip 15pt  6.7    \hskip 15pt     
       \hskip 20pt 0.65   \hskip 15pt  0.37        
       \hskip 18pt 3.5    \hskip 26pt 0.55 $

\ni
\vskip.1truein
\hrule  
\vskip.1truein

\vfill \eject

\baselineskip=24pt

\centerline {FIGURE CAPTIONS}

\bsni 
Fig. 1. Schematic equilibrium phase diagram (concentration of B-atoms, $c$,
versus temperature, $T$) for the model A-B alloy.  The circles show the
conditions at which our computer experiments were performed.

\bsni 
Fig. 2. Plots of configurations of the local order parameter, $\eta_{\bf
l}$, (by means of brightness $\eta_{\bf l} = -1$ being white and $\eta_{\bf
l} = +1$ being black), at three time points, $t^\thd =\ 5,\ 20,\ 40$ (time,
$t$ in units of $MCS$, from top to bottom) for each of three
concentrations, $c =0.15$ , 0.25, and 0.35.  The disordered phase has
$\eta_{\bf l} \approx 0$ and appears grey.  The lattice size is $128 \times
128$.

\bsni 
Fig. 3. Plot (by means of brightness, black meaning high values) of the
structure function, $S_{\bf k}$, in the entire Brillouin zone (top) and in
its parts related to decomposition (bottom, with higher contrast) at
$t^\thd = 5$ (time, $t$ is given in units of $MCS$), for the concentration
$c =0.25$.

\bsni 
Fig. 4. (a) Plots of the spherically averaged structure function for
ordering, $S_\pi(k)$, at $t^\thd =\ 2,\ 5,\ 10$ (time, $t$ is given in
units of $MCS$) for concentration $c =0.25$. (b) $S_0(k)$ at times $t^{1/3}
= 5, 10, 20$.

\bsni 
Fig. 5.  Dependence of the characteristic wavelength for (a) ordering,
$\lambda_\pi$, and (b) decomposition, $\lambda_0$, on time, $t$ (in units
of $MCS$) at three concentrations: $c = 0.15$ (diamonds), 0.25 (stars) and
0.35 (triangles).

\bsni 
Fig. 6 (a) Scaling of the structure function spherical average for
ordering, $F_\pi(k/k_\pi)$ where $k_\pi = a^\prime _\pi(t^{1 \over 3} +
b_\pi)$, for $c = 0.15, 0.25$ and $0.35$.  The full line is a fit with
eq.(22).  (b) Scaling of the structure function spherical average for
decomposition, $F_0(k/k_o)$, where $k_0 = {\acute a}_0(t^{1 \over 3} +
b_0)$ (at the same three concentrations).  The full line is a fit with
eq. (21), with the parameters given in Table 2.
\bsni
Fig. 7. Full-width at half maximum of the scaling function for
decomposition normalized according to eq.(19), as a function of the volume
fraction $f$ of the B-rich phase (crosses).  Data obtained with a
two-dimensional Ising model with attractive nearest neighbor interaction
between like atoms (taken from [9]) are also shown for comparison
(diamonds).

\magnification=1200
\baselineskip=24pt
\def\ni{\noindent}
\def\ssni{\smallskip\noindent}

\end